\documentclass[psfig,11pt]{article}
\usepackage{graphics}
\newcommand {\msol}{\mbox{M$_{\odot}$}}
\newcommand{\mhtwo} {\hbox{$M_{{\rm H}_2}$}}

\newcommand{\ratioo} {N({\rm H}_2) / I_{\rm CO}}
\newcommand{\kms}   {{\rm \, km \, s^{-1}}}
\newcommand{\K}     {{\rm \, K}}
\def \ergs{{\rm \, erg \, s^{-1}}}
\def \MHI{M_{\rm HI}}
\def \Mo{{\rm \, M_\odot}}
\def\ga{\lower.5ex\hbox{$\; \buildrel > \over \sim \;$}}
\def\la{\lower.5ex\hbox{$\; \buildrel < \over \sim \;$}}
\begin{document}

{\Large Formation of Molecular Gas in the debris of violent 
Galaxy Interactions} \\

Jonathan~Braine, Observatoire de Bordeaux, UMR 5804, CNRS/INSU, B.P. 89, 
  F-33270 Floirac, France \\

Ute Lisenfeld, Institut de Radioastronomie Millim\'etrique, Avenida 
Divina Pastora 7, NC18012 Granada, Spain \\

Pierre-Alain Duc, Institute of Astronomy, Madingley Rd., Cambridge, CB30HA, 
UK and \\
CNRS and CEA/DSM/DAPNIA Service d'astrophysique, Saclay,
91191 Gif sur Yvette cedex, France\\

St\'ephane Leon, ASIAA, Academia Sinica, P.O. Box 1-87, Nanking, Taipei 115, Taiwan

\bigskip\bigskip

\baselineskip=7mm

{\bf In many gravitational interactions between galaxies, gas and stars that
have been torn from  either or both of the precursor galaxies can collect in
'tidal tails'. Star formation begins anew in these regions to produce 'tidal
dwarf galaxies' \cite{Mirabel92} \cite{Duc94a} \cite{Duc97b} 
\cite{Duc98b}, giving insight into the process of galaxy formation through
the well-defined timescale of the interaction. 
But tracking the star formation process has proved to be
difficult: the tidal dwarf galaxies with young stars showed no evidence of the
molecular gas out of which new stars form \cite{Brouillet92} \cite{Walter99}
\cite{Smith94} \cite{Smith99}.  Here we report the
discovery of molecular gas (carbon monoxide emission) in two tidal dwarf
galaxies. In both cases, the molecular gas peaks at the same location as the
maximum in atomic-hydrogen density, unlike most gas-rich  galaxies. We infer
from this that the molecular gas formed from the HI, rather than being torn
in molecular form from the interacting galaxies. Star formation in the tidal
dwarfs appears to mimic that process in normal spiral galaxies
like our own.}

\bigskip

Tidal Dwarf Galaxies (TDGs) 
are  gas-rich irregular galaxies  made out of stellar and gaseous 
material pulled out by tidal forces from the disks of the colliding 
parent galaxies into the intergalactic medium \cite{Zwicky56}  
\cite{Schweizer78} \cite{Hibbard96} \cite{Ducrev}.
They are found at the ends of long tidal tails, sometimes  
100 kpc from the nuclei of their progenitors, and host active 
star-forming regions.  TDGs contain two main stellar components:
young stars recently formed by collapse of expelled atomic 
hydrogen (HI) clouds, and an older 
stellar population, at least 1 Gyr old, originally part of the disk of
the parent galaxies.  Their overall gaseous and stellar
properties range between those of classical  dwarf irregular and blue compact
dwarf galaxies, with the exception of their metallicity which is higher --
typical of the outer disk of a spiral \cite{Ducrev}.  
Whether a large fraction of dwarf galaxies were formed through tidal 
encounters in the early universe when spiral galaxies were more 
gaseous and less metal rich and collisions more frequent  is an open 
question and one of the drivers to study 
TDGs. One way  to answer this question  is the dark matter content
of dwarf galaxies. Observations of ordinary dwarf 
galaxies show that a lot of dark matter, or mass that is in some sotofar 
invisible form, is necessary to  account for their rotation 
velocities.   Numerical simulations of gravitational
interactions indicate that TDGs should have very little dark matter 
\cite{Barnes92} if the dark matter is, as currently believed, in form of 
a large halo and not in, say, a rotating disk.
Thus, {\it if} TDGs are found to possess the same dark matter 
properties as other dwarf galaxies then powerful constraints are
placed on the form of dark matter.
If TDGs do {\it not} contain dark matter as ordinary dwarf galaxies,
then tidal interactions cannot be the principal formation mechanism
for these small galaxies nor can dark matter be part of galactic disks. 


The observations were carried out with the 30meter telescope operated by the
Institut de Radioastronomie Millim\'etrique (IRAM) on Pico Veleta, Spain
in June  of 1999.  
Carbon Monoxide (CO) emission is detected in the Southern TDG in Arp~105 
(Figure 1 ; hereafter A105S) and 
main TDG in Arp~245 (Figure 2; hereafter A245N)
in both the ground state CO($J=1-0$) and 
the CO($J=2-1$) transitions.  Small maps were made of both sources
to localise the CO emission with respect to the atomic hydrogen
(HI), ionized gas (H$\alpha$), and optical continuum  \cite{Duc97b}
\cite{Duc99b}.  
The central (0,0) CO(1--0) spectra are shown in Fig. 3 along
with the HI spectra at the same positions with a similar beamsize.
The CO(1--0) luminosities and derived H$_2$ mass estimates (see Table 1)
of A105S and A245N are far above those of other dwarf galaxies 
\cite{Taylor98}.  Despite the different environments,
the star formation efficiency, defined as the rate of star formation
per mass of molecular gas, is quite close to that observed in the 
Milky Way and other spiral galaxies \cite{Kennicutt98}.

Small CO maps have been made consisting of six positions
towards A105S and four positions towards A245N. In both cases,
the CO peaks at the HI column density maximum and the dynamics of 
the atomic and molecular components are virtually identical (Figure 3).
In spiral galaxies, on the other hand, HI and 
CO have very different distributions (see {\it e.g.} \cite{Guelin93}
\cite{Braine4414a}), showing that the molecular gas 
that we have found in the TDGs has not simply been torn off the parent 
galaxies together with the HI but has formed {\it in situ}.
Although the calculations 
were performed for post-shock gas, an estimate of the molecule 
formation time is $t \sim n^{-1}$Gyr \cite{Hollenbach89} where $n$ 
is the density of the atomic medium in particles cm$^{-3}$.
Numerical simulations of Arp~245 \cite{Duc99b} yield an age of  
about 100 Myr for A245N and a
rough age estimate for A105S can be obtained by dividing the 
projected distance to the spiral by the relative radial velocity, 
yielding about 200 Myr \cite{Duc97b}, sufficient for H$_2$ formation 
in standard atomic hydrogen clouds ($\overline{n} \sim 10$cm$^{-3}$).
The dust on which the H$_2$ forms is captured from the parent
galaxies and present in the atomic gas \cite{Neininger96} \cite{Dumke97}
\cite{Braine3079}.  {\it The molecular gas has formed 
inside the HI clouds and star formation is proceeding in a standard 
way from the molecular gas.}

Our observations show that the molecular gas is an important 
component in the visible mass budget of TDGs, between $\sim 20$\% and
$\ga 50$\% of the atomic hydrogen mass (see Table 1).
The fact that we detect large
quantities of molecular gas and that we have every reason to 
believe that this gas is the result of conversion from HI into 
H$_2$ indicates that the central regions of these objects should be
gravitationally bound.  If the HI were dense enough pre-encounter then
CO would form and be routinely detected beyond R$_{25}$ (the optical 
radius) in galactic disks, like HI -- it is not \cite{Guelin93}
\cite{Neininger96} \cite{Dumke97} \cite{sage}.  
While it was clear that TDGs are kinematically decoupled from their
parent galaxies, prior to these data, the evidence that TDGs were
bound was morphological -- the accumulation of matter at the tips
of the tidal tails and the presence of star forming regions.  
Although higher angular resolution is necessary,
this conclusion provides firmer ground for the calculation of the 
dynamical mass, which relies on the assumption 
that the object is gravitationally bound and in equilibrium, and
thus for the determination of dark matter which
by definition is detected as  discrepancy between the velocities expected 
based on the mass of what we see directly and those observed. 

\bibliographystyle{unsrt}
\bibliography{tdg_nature}

\newpage

\begin{table}
\begin{center}
\begin{tabular}{lll}
 & A105S & A245N \\
\hline
\smallskip 
RA(J2000) & 11 11 13.5 & 09 45 44.1 \\
\smallskip 
Dec(J2000)& +28 41 20 & -14 17 28 \\
\smallskip 
HI velocity (LSR)& $cz=8890 \kms$ & $cz=2175 \kms$ \\
\smallskip 
adopted distance &  115 Mpc & 31 Mpc \\
\smallskip 
$L_{\rm H\alpha}$ & $1 - 2 \times 10^{40}\ergs$ & $7 \times 10^{39}\ergs$ \\
\smallskip 
M$_{\rm B}$, L$_{\rm B}$/L$_{\rm B}\odot$ &-16.9, $9 \times 10^8$
 & -17.25, $1.2 \times 10^9$ \\
\smallskip 
B -- V & 0.3 & 0.55 \\
\smallskip 
$\MHI$ & $5 \times 10^8\Mo$ & $9 \times 10^8\Mo$\\
\hline
\smallskip 
\mhtwo (\msol) & $\ge 2.2 \times 10^8$ & $\ge 1.4 \times 10^8$ \\
\end{tabular}
\caption[]{{\baselineskip=7mm
Properties of the Arp~105 and Arp~245 Tidal Dwarf Galaxies.
The data are from 
\cite{Duc94a} \cite{Duc97b} for A105S and \cite{Duc99b} for A245N.
Position is (0,0) of CO map and velocity is zero of spectra (Fig. 3).
M$_{\rm B}$ and L$_{\rm B}$ include
a correction for galactic absorption of 0.3 magnitudes for A245N.  
A105S is at high galactic latitude so no correction is applied.
The molecular gas mass is estimated using a $\ratioo$ factor 
of 2 $\times 10^{20} \K\kms$cm$^{-2}$ and likely represents a lower limit 
because weaker, undetected, CO emission may be present at other positions.   
We have included the mass of Helium in the molecular clouds.
Relative to the velocities of the TDGs, the spiral and elliptical in the 
Arp~105 system have velocities of -130 and -400 km/s.  In Arp~245, the
velocities of the spirals NGC 2992 and NGC 2993 are 155 and 245 km/s 
with respect to the TDG.
}}
\end{center}
\end{table}

\begin{table}
\begin{center}
\begin{tabular}{llllll}
\hline
Source & offset & I$_{\rm CO}$ & rms &  vel. & $\Delta$V$_{\rm fwhm}$ \\
\smallskip 
 & ($\delta$RA,$\delta$Dec) & K km/s & mK & km/s & km/s\\
\hline
A105S & (0,0) & 0.3$\pm.05$ & 2.5 & 19$\pm6$ &38$\pm10$ \\
\smallskip 
 & & 0.2$\pm.05$ & 3.5 &16$\pm5$ & 25$\pm10$ \\
\smallskip 
A105S & (10,0) &0.1$\pm.05$ & 3 & 59$\pm6$ & 19$\pm10$ \\
\smallskip 
A105S & (-10,0) &0.15$\pm.05$ & 2.8  & -4$\pm5$ & 22$\pm13$ \\
\smallskip 
A105S & other &0.15$\pm.05$ & 2.5 & 15$\pm6$ & 15$\pm8$ \\
\smallskip 
A245N & (0,0) & $1.3\pm.1$ & 4.9 &-32$\pm4$ &48$\pm10$ \\
\smallskip 
& & $2.0\pm.5$& 18  &-32$\pm10$ &66$\pm20$  \\
\smallskip 
A245N & (3,14) & $0.6\pm.15$ & 7 &-33$\pm10$ &47$\pm15$ \\
\smallskip 
A245N & (0,-10) & $0.9\pm.15$ & 7.6 &-27$\pm7$ &44$\pm14$  \\
\smallskip 
A245N & (0,-20) & $0.5\pm.2$ & 9.5 &-27$\pm12$ &32$\pm15$  \\
\end{tabular}
\caption[]{
{\baselineskip=7mm 
Molecular gas in Tidal Dwarf Galaxies.
The CO observations presented here provide the first detections 
of molecular gas in TDGs.
The offset is in arcseconds with respect to the position given 
in Table 1 and the red circle in Figures 1 and 2.  I$_{\rm CO}$
is the flux of the CO line expressed in Kelvin km/s.
The lines without source and offset are the CO(2--1) 
observations of the preceding source and position.  Noise levels (rms) 
are given for channel widths of 2~MHz in the CO(1--0) line and
2.5~MHz in the CO(2--1) line.  
The velocity of the line center is with respect to the HI velocity 
given in Table 1.  Note that the ``detections''
of the off-center A105S positions are uncertain so the velocities
and line widths may be meaningless.  
The spectra for each position were averaged, a continuum level was 
then subtracted such that the average flux outside the line
window is zero, and resulting spectra were smoothed to yield the
results presented in Fig. 3 and Table 2.
No baselines other than the continuum level are subtracted from the data.
The A105S ``other'' position
represents the average of the spectra for the (0,5) (0,-5) and (0,-10)
positions. Taken individually, these points were not detected and 
yield 3$\sigma$ limits of I$_{\rm CO} \la 0.3$K km/s. 
The CO emission from A105S is consistent with a near-punctual
source, much like for the optical and HI.  
The angular resolutions are respectively 
22$''$ and 11$''$; beam efficiencies are 0.72 and 0.48.
}} 
\end{center}
\end{table}

\clearpage

\pagebreak

\newpage

\begin{figure}
\caption {{\baselineskip=7mm
The Southern Tidal Dwarf Galaxy (blow-up) in the interacting system 
Arp~105 (NGC 3561 \cite{arpatlas}; ``the Guitar'').  
HI emission contours are superposed on a blow-up 
of a V band image of A105S \cite{Duc97b}.
The frame is 4.4$'$ x 5.9$'$; North is up and East to the left.
Red circle is (0,0) position of CO observations
and represents the FWHM (22$''$) of the CO(1--0) beam.
The Arp~105 system \cite{Duc94a} \cite{Duc97b}
is an interaction between a spiral and an elliptical
which has generated an HI-rich extended TDG at the end of the Northern
tidal tail and a more compact TDG at the tip of the Southern tail
from the spiral.  Arp~105~South (A105S) 
contains roughly 5 $10^8$ \msol\ of HI and strong H$\alpha$ emission,
corresponding to a star formation rate of $\sim 0.2$\msol/yr.
Nonetheless, stellar synthesis models \cite{Fritze98}
of A105S indicate that the stellar mass is dominated by the old 
spiral disk population while the luminosity comes in majority from stars 
formed {\it in situ}. 
 }}
\end{figure}

\begin{figure}
\caption {{\baselineskip=7mm
The Tidal Dwarf Galaxy (blow-up) in the interacting system 
Arp~245 (NGC~2992/3 \cite{arpatlas}).  
HI emission contours are superposed on a 
blow-up of a V band image of A105S  \cite{Duc99b}. 
The frame is 5.8$'$ x 7.4$'$; North is up and East to the left.
Red circle is (0,0) position of CO observations
and represents the FWHM (22$''$) of the CO(1--0) beam.
Arp~245 is an interaction between two spirals.  The TDG, A245N,
has been formed in the tidal tail which stems from NGC~2992 and
contains nearly twice as much HI as A105S but slightly weaker
H$\alpha$ emission.  The old stellar population is more prominent 
in A245N than in A105S \cite{Duc99b}.
The physical size and total HI mass of Arp~245 are smaller than 
in Arp~105.} }
\end{figure}

\begin{figure}
\caption {{\baselineskip=7mm
CO(1--0) and HI spectra of the (0,0) position of A105S (left) and
A245N (right). 
The velocities and line widths of the CO and HI emission are very similar.
Towards the very compact TDG A105S, the CO emission is not resolved. 
The CO emission in A245N is extended with detections in {\it at least}
3 of 4 observed points.  The H$\alpha$ emission in A245N \cite{Duc99b}
decreases substantially towards the (0,-10) offset position while 
the HI \cite{Duc99b} and CO (see Table 2) are still strong.
In contrast, the H$\alpha$ emission towards the (3,14) position is
comparable to the center whereas the CO and HI have decreased 
significantly.
The temperature scale (main beam) is in milliKelvins and 
velocities in km/s. HI intensity is in arbitrary units.} }
\end{figure}

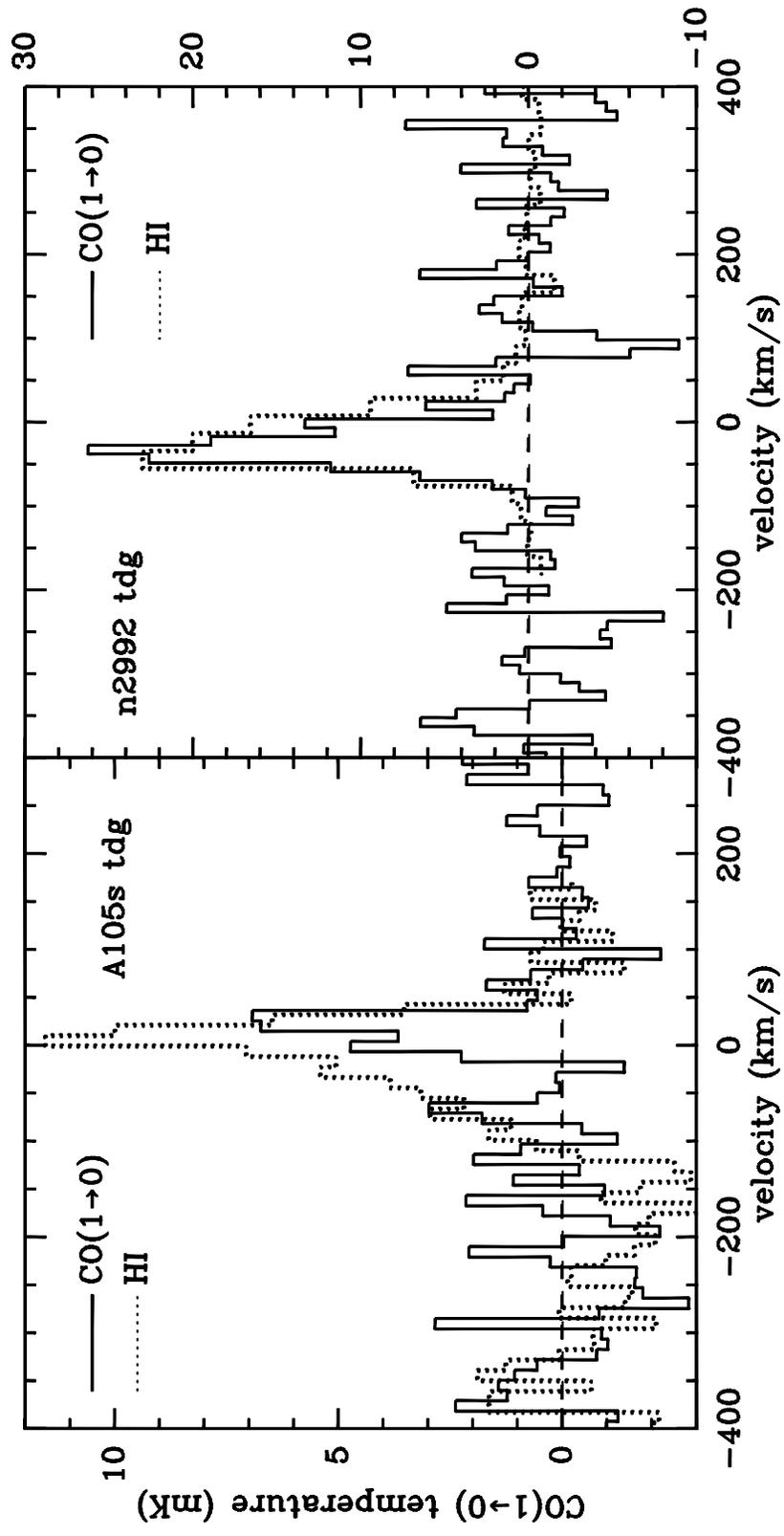
\end{document}